\documentclass[12pt]{iopart}
\usepackage{iopams}
\usepackage{graphicx}

\def\Journal#1#2#3#4{{\it {#1}} {\bf#2} #3 (#4)}
\def\PRL{Phys. Rev. Lett.}
\def\PRC{Phys. Rev.}

\def\JPG{J. Phys. G}
\begin{document}

\title[$ \Sigma (1385)$ Results and Status of the $\Theta ^{+}$ in STAR]{$ \Sigma (1385)$ Results and Status of the $\Theta ^{+}$ in STAR}

\author{Sevil Salur\footnote[1]{sevil.salur@yale.edu} for STAR Collaboration\footnote[2]{For the full STAR Collaboration author list, see appendix 'Collaborations' of this volume.} }

\address{ Physics Department, Yale University, \\Sloane Physics Laboratory, P.O. Box 208120,\\ New Haven, CT, 06520-8120, USA}

\begin{abstract}
The $\Sigma (1385)$ analysis and the current status of pentaquark
search with the STAR detector are reported. The corrected $p_{T}$ spectra and the yields of
the $\Sigma ^{\pm} (1385)$ and their antiparticles in the most
central Au+Au as well as elementary p+p collisions are presented. A
comparison of the $\langle p_{T} \rangle$ of observed particles
suggests a similar behavior for particles with mass greater than 1.2 GeV in
p+p and Au+Au collision environments. Acceptance and efficiency
studies with simulations show that the (anti)pentaquarks should be
found at the 3 \% level.
\end{abstract}




\section{Introduction}
The observation of a five-quark bound system consisting of
uudd$\overline{s}$, referred to as the $\Theta^{+}$ pentaquark, have been reported in
photon-nucleus and kaon-nucleus reactions
\cite{{leps},{clas},{diana}}. The presence of this state was
predicted by R. L. Jaffe with multiquark bag models
\cite{{jaffe1},{jaffe2}} and later by D. Diakonov et al. using
chiral soliton models of baryons \cite{diakonov}. The high
energies and particle densities resulting from
collisions at the Relativistic Heavy Ion Collider (RHIC) are
expected to be ideal environments for pentaquark
production \cite{liewen,Ko,raf:1,Randup}.

During the expansion of the hot and dense matter (fireball) created in Au+Au collisions, chemical freeze-out is
reached when the hadrons stop interacting inelastically. Elastic
interactions continue until thermal freeze-out. Due to the very
short lifetime ($\tau < \tau_{fireball}$) of resonances, a large
fraction of the decays occur during this time. The elastic
interactions of decay products with the surrounding particles
result in a signal loss in the particle identification though this is offset by
secondary interactions which increase the resonance yield (e.g.
$\Lambda$ + $\pi$ $\rightarrow$ $\Sigma (1385)$, K+p $\rightarrow$
$\Theta^{+}$). The contribution of re-scattering and regeneration
on the total observed yields depends on the time span between the
chemical and thermal freeze-out and the lifetime of each
resonances \cite{tor01,ble02}. Thus the study of resonances
provides an additional tool in the determination of the hadronic
expansion time between chemical and thermal freeze-out by
comparing resonance to stable particle ratios. The data analyzed
were taken by the STAR (Solenoidal Tracker At RHIC)
experiment\cite{stardet}, one of the four experiments at RHIC. The
large acceptance of STAR's Time Projection Chamber (TPC) is ideal for
such rare particle searches.

\section{Particle Identification}

The STAR trigger detectors consist of two zero degree calorimeters (ZDCs) which are situated $\sim$ 17 m
downstream of the nominal interaction point as well as a central trigger barrel (CTB) surrounding the TPC.
A minimum bias trigger is defined by the coincident measurement of spectator neutrons in both ZDCs. The CTB
an array of scintillator slats, is used to detect event multiplicity, is used to trigger on the 10\% most
central Au+Au collisions. For p+p collisions, a minimum bias trigger is defined by coincidences in the
two beam-beam counters, which are again scintillator detectors and are situated around the beam pipe,
approximately 2 m from the center of the TPC. The main detector component of STAR is the TPC, which, together
with the magnetic field information, is used to identify stable charged particles via energy loss per unit
length and long-lived weakly decaying ($c \tau \sim $ few cm) neutral particles such as $\Lambda$ and $K^{0}_{S}$
via their decay topology. The direct measurement of resonances is not possible due to their short lifetimes
($c\tau_{\Sigma (1385)}=5 \;fm$) \cite{rho,kstar}. Instead, $\Sigma (1385)$ resonances are identified by
the invariant mass after combining $\pi$ with $\Lambda$ decay particle candidates. A mixed event
technique, where particles from different events are reconstructed using the same technique, is
used to determine the background for uncorrelated pair combinations. The $\Sigma (1385)$ signal
is obtained by the subtraction of this normalized mixed event background from the
invariant mass distribution \cite{sal01}. This technique can
directly be applied for the search and possible identification of
pentaquarks. For the case of the $\Theta^{+}$ pentaquark study the
invariant mass reconstruction is via $\Theta ^{+} \rightarrow
K^{0}_{S}+p$, and subtraction of the normalized mixed event
background is used \cite{sal02}.

\section{The $\Sigma (1385)$ Analysis}

\begin{figure}[t!]
  \centering
  \includegraphics[width=0.9\textwidth]{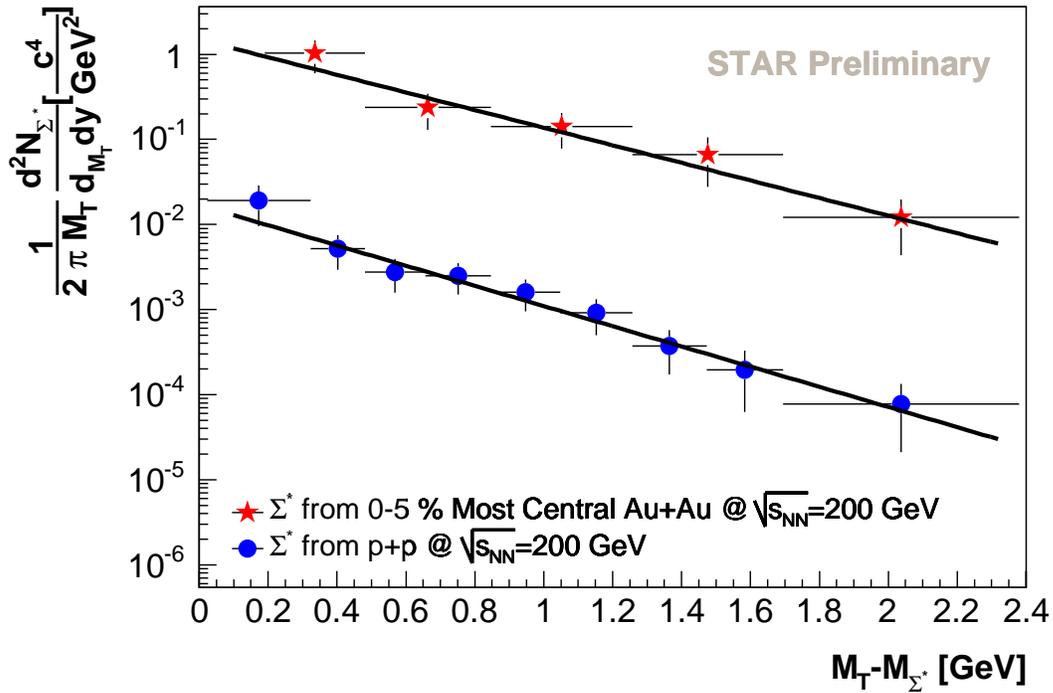}
  \caption{The transverse mass spectrum
for $\Sigma (1385)$ drawn as stars for the
$0-5\%$ most central Au+Au and circles for p+p collisions at $\sqrt{s_{NN}}=200$ GeV.
Blue circles are for p+p and red stars are for Au+Au collisions.}
  \label{mtspectra}
\end{figure}

The transverse mass ($m_{T}=\sqrt{p_{T}^{2}+m^{2}}$) spectra of
$\Sigma ^{\pm} (1385)$ and their antiparticles in p+p (circles)
and Au+Au (stars) collisions are shown in Fig.~\ref{mtspectra}.
The spectra are corrected for acceptance and efficiency by
embedding Monte-Carlo simulated resonances into real p+p and Au+Au
events. The solid lines in Fig.~\ref{mtspectra} represent
exponential fits to the data with the function directly
proportional to the yield (dN/dy) and inversely proportional to
temperature ($T^{2}+m_{0}T$).  The data coverage is 91\% in p+p and
85\% in Au+Au collisions of the fully integrated yield. The $\langle p_{T}
\rangle$ is derived from the full range integration of the
corresponding exponential fit. Table~\ref{table1} presents the inverse slope parameter (T), the
$\langle p_{T}\rangle $ and the yield (dN/dy) for the summed signal of $\Sigma^{\pm}(1385)$ together with
their antiparticles, for both p+p and Au+Au collisions at $\sqrt{s_{NN}}=200$ GeV.

\begin{table}[b!]
\begin{tabular}{|l|c|c|l|} \hline
  & T [MeV] & $\langle$p$_{\rm T}$$\rangle$ [GeV/c] & Yields (dN/dy)   \\
  \hline \hline
$\Sigma^{\pm} (1385)$ in p+p & $358 \pm 47$  & $1.08 \pm 0.15$ & $(4.66 \pm 0.98 ) \times 10^{-3}$  \\
\hline $\Sigma^{\pm} (1385)$ in Au+Au & $420 \pm 84$  & $1.20 \pm 0.24$ & $4.72 \pm 1.38 $ \\
\hline
\end{tabular}
\centering \caption{Temperature T, $\langle p_{T}\rangle $ and
yield obtained from the exponential fits of \\ the $p_{T}$ spectra
in Fig.~\ref{mtspectra} for elementary p+p and 0-5\% most central
collisions. \\The statistical uncertainties are given and the
systematical error due to norm-\\alization of the background,
$\sim 15\%$, has to be included in the given values.}
\label{table1}

\end{table}

\begin{figure}[t!]
  \begin{center}
  \includegraphics[width=0.9\textwidth]{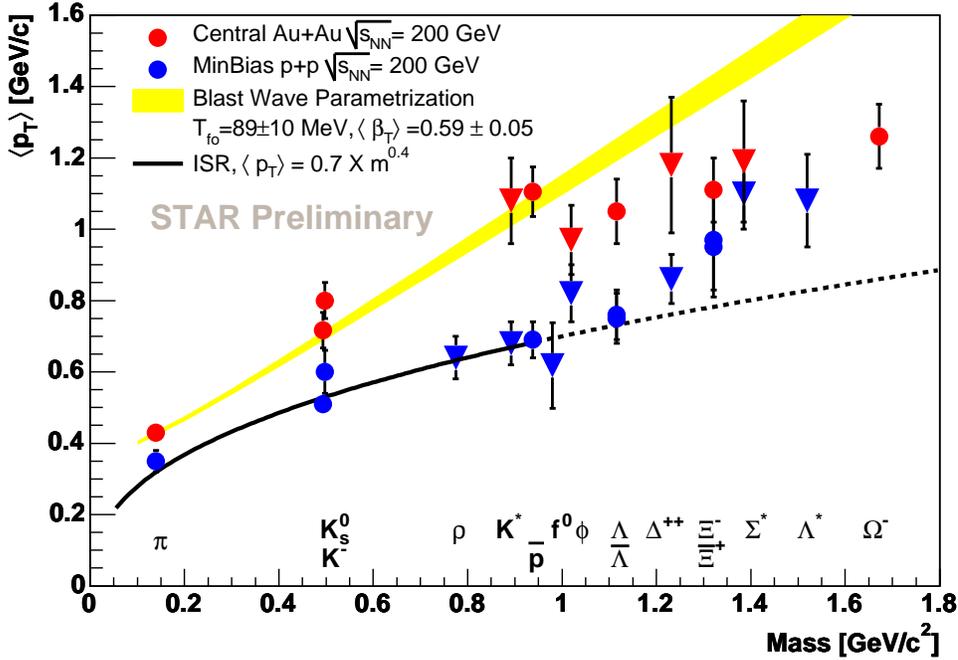}
\end{center}
\caption{The $\langle p_{T} \rangle$  vs particle mass measured in
p+p and Au+Au collisions at $\sqrt{s_{NN}}=$ 200 GeV. While the
circles represent the stable particles, triangles represent the
resonances.  The black curve represents the ISR parametrization
from $\pi$, K and p for $\sqrt{s_{NN}}=$ 25 GeV p+p collisions.
The shaded band is the blast wave fit using $\pi$, K and p for
Au+Au collisions.}
  \label{meanptvsmass}
\end{figure}

The $\langle p_{T} \rangle$  for various particles in both p+p
and Au+Au collisions at $\sqrt{s_{NN}}=$ 200 GeV as a function of their mass are presented in
Fig.~\ref{meanptvsmass}. The behavior of $\langle p_{T}\rangle$
vs. mass for the various particles in p+p and Au+Au collisions is
compared to two parameterizations. The black curve is an empirical fit
to the ISR $\pi$, K and p data \cite{ISR01} and the shaded band is a blast wave fit
using $\pi$, K and p data in Au+Au collision system \cite{mult130}. The empirical parametrization
for the ISR data at $\sqrt{s}=$ 25 GeV in p+p collisions, can
describe the behavior of the lower mass particles, such as $\pi$,
K and p, despite the fact that our collision energy is one order
of magnitude higher. However, this empirical parametrization does
not represent the behavior of the higher mass particles. Similarly
the blast wave parametrization which can describe the lower mass
particles in Au+Au collisions ($\sim 98\%$ of all the particles observed) fails to
explain the behavior of higher mass particles.

The comparison of lighter, weakly decaying particles and heavier resonances
in p+p and Au+Au collision environments shows a similar behavior of $\langle p_{T} \rangle$
for the higher mass particles. It is expected that resonances with
higher transverse momentum are more likely to be reconstructed in Au+Au collisions
because of their longer relative lifetimes due to Lorentz
contraction. This means they are more likely to decay outside
the medium and hence their daughter particles would interact less
with the medium. Any loss at low $p_{T}$ would increase the $T$
parameter of the $p_{T}$ spectra for the central Au+Au collisions
with respect to p+p collisions. However we do not see any
significant increase in the $T$ parameter for $\Sigma (1385)$ from
p+p to the most central Au+Au collisions within the statistical
and systematic errors. If higher mass particles are produced by different
production mechanisms than the lower mass particles, we might be
introducing a bias in our measurement in p+p
collisions. This may be due to a number of reasons which include higher mass
particles being produced in more violent (mini-jet) p+p collisions and/or the heavier particles
flow radially with a smaller velocity than the lighter mass
particles (such as $\pi$ mesons) in the most central Au+Au
collisions. The $\langle p_{T} \rangle$ measurement of the higher mass
resonance $\Sigma (1385)$ shows this merging behavior.
Similarly the study of radial flow of heavier particles can give
us valuable information about whether the heavier particles flow
less with respect to lower mass particles.
%\begin{figure}[h!]
%  \centering
%  \includegraphics[width=0.7\textwidth]{sigmaratio.eps}
%  \caption{The $\Sigma^{\pm}(1385)/\Lambda $  vs collision energy measured\\ in
%e+e, p+p and Au+Au collisions. }
%  \label{sigmaratio}
%\end{figure}

A comparison of $\Sigma (1385) / \Lambda$ ratios in p+p and $0-5\%$
central Au+Au collisions shows no suppression from p+p to Au+Au.
An observed suppression of $K^{*}(892)/K$ and $\Lambda ^{*}(1520)/\Lambda$
ratios in Au+Au collisions with respect to p+p collisions suggests
a signal loss via re-scattering in the medium by thermal models \cite{bec02,pbm01}.
As the lifetime of a $\Sigma (1385)$ is one third that of the $\Lambda ^{*}(1520)$, it
is expected that there should be an even greater signal loss via re-scattering. The
observed unsuppressed ratio requires a significant regeneration mechanism in order to
recover the signal loss via re-scattering. The measured $\Sigma ^{\pm}(1385)/\Lambda$
ratio is $0.295\pm0.086$ for the 0-5\% most central collisions which is about a factor of
2 below the microscopic model UrQMD predictions at $\sqrt{s_{NN}}=200$ GeV.  This suggests
that the assumed regeneration cross-section, included in UrQMD calculations, is too high  \cite{ble04}.
Both UrQMD and thermal production models should be revised
in light of the resonance measurements.

\section{Status of Current $\Theta ^{+}$ Studies}

\subsection{Monte Carlo Studies}

\begin{figure}[t!]
  \centering
  \includegraphics[width=0.8\textwidth]{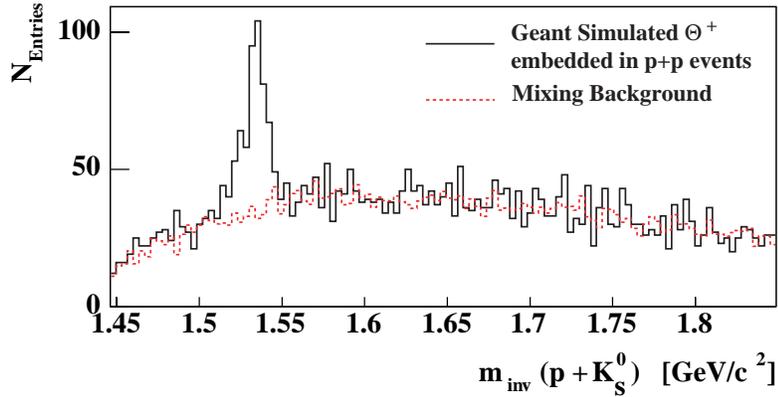}\\
  \caption{Invariant mass spectrum of the Monte Carlo simulated $\Theta^{+}$ embedded in real p+p events
  with the mixing technique. Black solid histogram is the signal and red dashed histogram is the mixed event background.
  The simulated signal can be clearly seen for 1 simulated particle per event.}
  \label{invmassnugget}
\end{figure}

To study the decay mechanism and optimize the applied cuts,
Monte Carlo simulations are used.   In this study, one Monte Carlo
$\Theta^{+}$ pentaquark is chosen from a thermal exponential
distribution with T = 250 MeV in the rapidity interval $\mid y\mid <1.5$ and
after a full TPC simulation is embedded into a single, real p+p
event. The chosen input width, 10 MeV/$c^{2}$, and the input mass,
1.54 GeV/$c^{2}$,
%\begin{figure}[h!]
%\centering
%  \includegraphics[height=5cm]{MC_input_theta.eps}\\
 %\includegraphics[width=0.7\textwidth]{simulation_2.eps}\\
  %\caption{Invariant mass spectrum of the Monte Carlo generated input
  %for\\ the $\Theta^{+}$ and $p_{T}$ vs rapidity of the $\Theta^{+}$ that can be identified
%with the TPC.} \label{fig:simulation}
%\end{figure}
are consistent with the observed mass and width of $\Theta^{+}$
\cite{{leps},{clas},{diana}}. %In Fig.~\ref{fig:simulation}, the
%invariant mass spectrum of simulated $\Theta^{+}$ and the TPC
%acceptance for the corresponding simulated particles is presented.
In Fig.~\ref{invmassnugget} the invariant mass spectrum of the
reconstructed $\Theta^{+}$ is presented. We find that $\sim3\%$ of
these Monte Carlo generated $\Theta^{+}$'s are successfully
reconstructed with this technique. The reconstructed width and the
mass is consistent with the Monte Carlo input.

\begin{figure}[t!]
\begin{minipage}[b]{0.5\linewidth} % A minipage that covers half the page
\centering
\includegraphics[width=8cm]{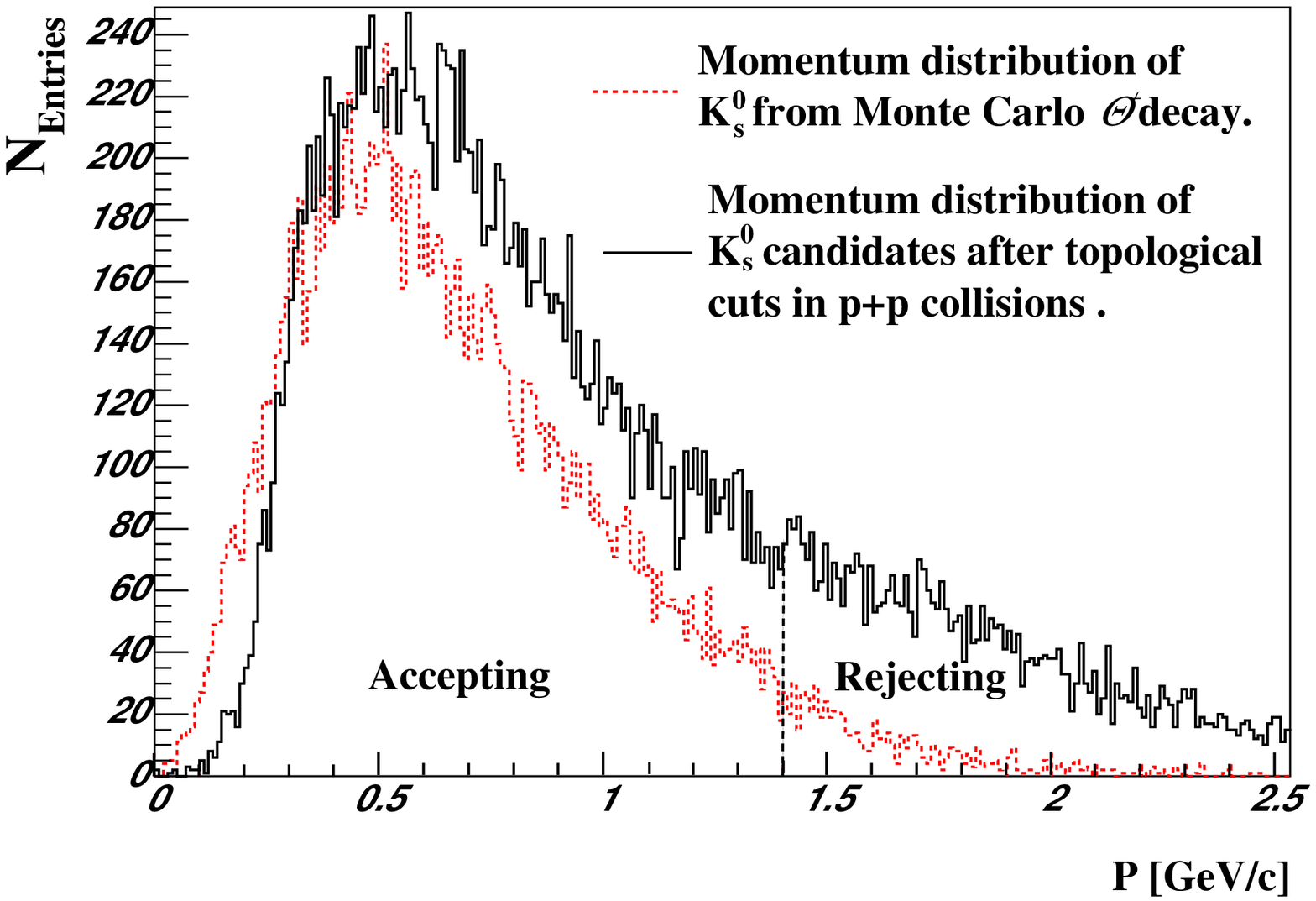}
\end{minipage}
\hspace{0.04cm} % To get a little bit of space between the figures
\begin{minipage}[b]{0.5\linewidth}
\centering
\includegraphics[width=8cm]{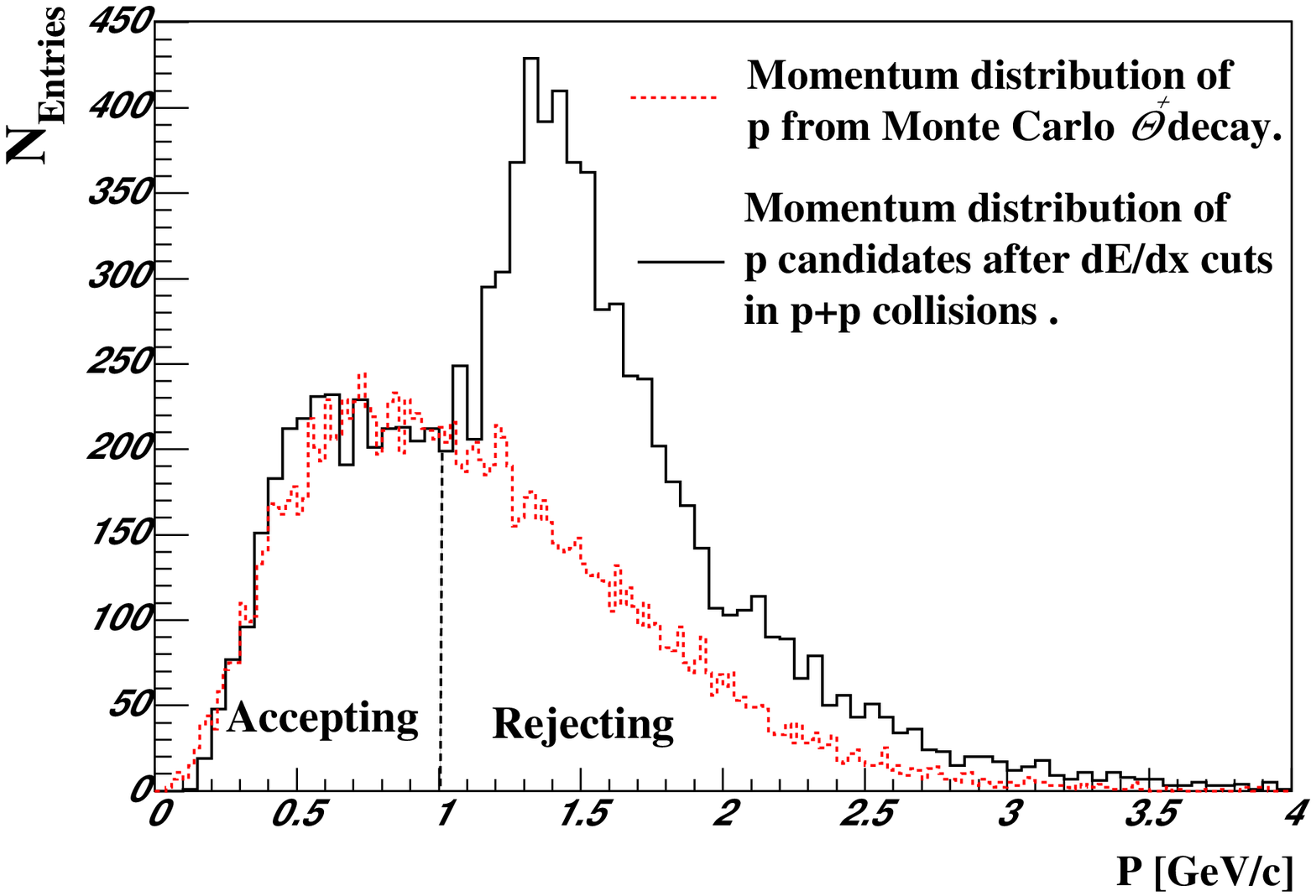}
\end{minipage}
 \caption{On the left is the $K_{S}^{0}$ momentum
distribution and on the right is the proton momentum distribution.
Black solid histograms is of the accepted $K_{S}^{0}$ on the left
and p on the right after the dE/dx cut and red dashed histograms
is of the decay daughters of the Monte Carlo generated  $\Theta
^{+}$ for the same number of events.} \label{momdis}
%\vspace{-0.2in}
\end{figure}

Using this technique, the decay properties with the simulated tracks such
as the momentum distribution of the decay daughters can be studied. In
Fig.~\ref{momdis}, the momentum distributions of the $K_{S}^{0}$ on
the left and proton on the right are presented. With this study we
can optimize momentum cuts to increase the signal-to-background
ratio. Clearly, an optimized cut to improve this ratio is to accept
protons with momentum less than 1~GeV/c while at the same time rejecting the ones
above this threshold. Further detailed studies on other variables
are needed to optimize the signal-to-background ratio.

\subsection{Feasibility Studies}
Assuming that the $\Theta^{+}$ production is $10-100\%$ of the $\Lambda
^{*}(1520) $ in p+p collisions, one can estimate the yield of the
$\Theta^{+}$. The preliminary $dN/dy$ of $\Lambda ^{*}(1520)$ at
mid-rapidity is 0.004 per event in p+p collisions
\cite{{ludo},{markert}}. There are 8 Million p+p events available
for this analysis and this corresponds to a production of $\sim30000$
$\Lambda^{*}(1520)$, giving a production range of $\sim3000$ to $\sim30000$
$\Theta^{+}$'s in these p+p events using the above estimates. As the efficiency of the
mixing technique is $\sim3\%$ and the branching ratio of the
$\Theta^{+}\rightarrow K_S^{0}+p$ is $\sim25\%$ (assuming that the
branching ratios of $\Theta\rightarrow K^{0}_{S}+p$ and
$\Theta^{+}\rightarrow K^{+}+N$ are each $50\%$), 20-200 of the
$\Theta^{+}$'s should be found. The background pairs per event in
the 1.54$\pm 5$ MeV mass range is 3200. This corresponds to a
significance  of between 0.25 and 3 \footnote[2]{The significance is defined as
$\frac{Signal}{\sqrt{2\times Background+Signal}}$}. Similarly one
can repeat the same study for Au+Au and d+Au collisions and
correspondingly predict a significance of 2-7 for 1.5 Million
Au+Au events and 1-16 for 10 Million d+Au for the predicted
production of one $\Theta^{+}$ per unit rapidity per collision
\cite{liewen,Ko,raf:1,Randup}. To estimate the yield for the d+Au
collisions we assume scaling with the number of participants ($N_{part}$). The mean number of
participants in d+Au is 8, in p+p it is 2, and in Au+Au it is 350
for the most central collisions. The lower limit is obtained from
p+p scaling while the upper limit is from Au+Au yield estimates.
The invariant mass spectra that are observed for the $\Theta^{+}$ in p+p,
d+Au and Au+Au collision events are consistent with
our estimations for the significance of the signal given our
current statistics.  The signal-to-background ratio depends highly on
the selection of events and applied cuts. To improve cuts and
understand the decay mechanism more detailed simulation
studies must be undertaken.
\section{Conclusions}
$\Sigma^{\pm}(1385)$ resonances and their antiparticles can be
reconstructed via event mixing techniques in p+p, d+Au and Au+Au
collision environments with STAR. The $\Sigma (1385) $  $\langle
p_{T} \rangle$ in p+p collisions is similar within the errors to
that measured in Au+Au collisions. This follows the behavior of
the other high mass particles ($m>1.2$ GeV) whose $\langle p_{T}
\rangle$ measurement in p+p approaches the Au+Au value. This
behavior suggests a possible smaller radial flow for heavy
particles with respect to $\pi$ and/or a more violent production
mechanism for heavy particles in p+p collisions.

 There is no significant suppression observed in
the $\Sigma (1385) / \Lambda$ ratios from  p+p to $0-5\%$ central
Au+Au collisions. Since the previously measured suppression of
$K^{*}(892)/K$ and $\Lambda ^{*}(1520)/\Lambda$ ratios agrees with a signal
loss via rescattering in the medium, the absence of suppression in the
$\Sigma (1385)$ ratio suggests a significant regeneration
mechanism to recover the signal lost via rescattering
\cite{ble03,tor01a}.

Acceptance and efficiency studies show that we should be able to
find $\Theta ^{+}$ pentaquarks at the few \% level with the current data-set
in d+Au and Au+Au collisions. Optimization of cuts to improve the
signal over background is in progress. There is a possibility of
measuring anti-pentaquarks at RHIC since the antibaryon to baryon
ratio is approaching to one \cite{ratio}. An upper limit to the
yields and production mechanisms of pentaquarks in Au+Au
collisions will be established in the 2004 run, which has a
70-fold increase in statistics over the current data.

\end{document}